\documentclass[10pt,twocolumn, conference]{IEEEtran}

\usepackage{amsmath}
\usepackage{amssymb}
\usepackage[dvips]{graphicx}
\usepackage{setspace}
\usepackage{epsfig}
\usepackage{epsfig}
\usepackage{srcltx}
\usepackage{srctex}

\newcommand{\figsize}{0.43}

\begin{document}
\title{Ergodic Capacity Analysis in Cognitive Radio Systems under Channel Uncertainty}
\author{\authorblockN{Sami Akin and
Mustafa Cenk Gursoy}
\authorblockA{Department of Electrical Engineering\\
University of Nebraska-Lincoln\\ Lincoln, NE 68588\\ Email:
samiakin@huskers.unl.edu, gursoy@engr.unl.edu}}
\date{}

\maketitle

\begin{abstract}
In this paper, pilot-symbol-assisted transmission in cognitive radio systems over time selective flat fading channels is studied. It is assumed that causal and noncausal Wiener filter estimators are used at the secondary receiver with the aid of training symbols to obtain the channel side information (CSI) under an interference power constraint. Cognitive radio model is described together with detection and false alarm probabilities determined by using a Neyman-Person detector for channel sensing. Subsequently, for both filters, the variances of estimate errors are calculated from the Doppler power spectrum of the channel, and achievable rate expressions are provided considering the scenarios which are results of channel sensing. Numerical results are obtained in Gauss-Markov modeled channels, and achievable rates obtained by using causal and noncausal filters are compared and it is shown that the difference is decreasing with increasing signal-to-noise ratio (SNR). Moreover, the optimal probability of detection and false alarm values are shown, and the tradeoff between these two parameters is discussed. Finally, optimal power distributions are provided.
\end{abstract}

\vspace{-0.1cm}
\section{Introduction}

The need for the efficient use of the scarce spectrum in wireless applications has led to significant interest in the analysis of cognitive radio systems which are proposed as systems that have the potential to operate in the licensed spectrum without disturbing the primary users and hence better utilize the spectrum \cite{cognitive}.
Since it is very important for the cognitive users not to interfere with the primary users during their transmission, detection of primary users' activity in the channel is the main concern in cognitive radio research. In \cite{qzhao} and \cite{yunxia} the authors developed an optimal strategy for opportunistic spectrum access by maximally utilizing spectrum opportunities in cognitive radio networks with multiple potential channels. Note that spectrum sensing brings along an essential decrease in the throughput of a system. Realizing this fact, the authors in \cite{Hoang} studied the tradeoff between channel sensing and throughput considering Shannon capacity as the throughput metric. Providing primary users with a sufficient protection against interference, they formulated an optimization problem and found out the optimal sensing time. Motivated by jointly detecting signal energy levels over multiple frequency bands rather than considering one band at a time, Quan \textit{et al.} in \cite{poor} introduced a novel wideband spectrum sensing technique.

An important characteristics of wireless communications is that channel conditions vary over time randomly due to mobility and changing environment  If the channel conditions are not known a priori, training sequences are frequently employed to perform channel estimation. Early studies conducted by Cavers \cite{cavers1} and \cite{cavers2} provided an analytical approach in pilot-assisted transmissions. In \cite{hassibi}, Hassibi and Hochwald  optimized the power and duration of training signals by maximizing a capacity lower bound in multiple-antenna Rayleigh block fading channels. Ohno and Giannakis \cite{giannakis} focused on a capacity lower bound and optimized the spacing of training symbols and training power by employing a noncausal Wiener filter for channel estimation at the receiver in slowly-varying channels. In their studies \cite{faycal}, Abou-Faycal \textit{et al.} employed a causal Kalman filter at the receiver in which all the past pilot symbols are used to estimate the channel coefficients. More recently, we in \cite{icc_akin} jointly optimized the pilot symbol period and power allocation among pilot and data symbols by maximizing the achievable rates in Gauss-Markov fading channels, and in \cite{spawc_akin} we used causal and noncausal Wiener filters at the receiver to estimate the channel coefficients.

In this paper, we study the training-based transmission schemes for cognitive radio systems in time-selective Rayleigh fading channels where cognitive radios are able to transmit data as long as they do not disturb the primary (licensed) users of the channel. We employ causal and noncausal Wiener filters at the receiver to obtain the estimates of the channel fading coefficients. We maximize a capacity lower bound by optimizing the training parameters. Note that even though the treatment is in general, we provide our numerical results for Gauss-Markov channel models.

\vspace{-0.2cm}
\section{Cognitive Channel Model and Channel Sensing}\label{sec:system}
In this paper, we consider a cognitive radio channel model communicating over a time-selective Rayleigh flat fading channel in which a secondary transmitter attempts to send information to a secondary receiver, probably interfering with the primary users. Initially, the secondary users perform channel sensing, and then depending on the primary users' activity, the secondary transmitter selects its transmission power, i.e., when the channel is busy, the average symbol power is $\overline{P}_{1}$, and when the channel is idle, the average symbol power is $\overline{P}_{2}$. For example, if $\overline{P}_{1}=0$, the secondary transmitter stops transmission when the primary users are sensed to be active. During the transmission, the discrete-time channel input-output relation in the $k^{\text{th}}$ symbol duration is given by
\begin{align}\label{input-out1}
&y_{k}=h_{k}x_{1,k}+n_{k}+s_{p,k}\quad k=1,2,\dots
\end{align}
if the primary users are in the channel. On the other hand, if the primary users are idle in the channel, we have
\begin{align}\label{input-out2}
&y_{k}=h_{k}x_{2,k}+n_{k}\quad k=1,2,\dots
\end{align}
In the above equations, $x_{1,k}$ and $x_{2,k}$ are the complex-valued channel inputs, and $y_{k}$ is the complex-valued channel output, and $\{n_{k}\}$ is assumed to be a sequence of independent and identically distributed (i.i.d.) zero mean Gaussian random variables with variance $\sigma_{n}^{2}$. In (\ref{input-out1}) and (\ref{input-out2}), $h_{k}$ denotes the fading coefficient between the secondary transmitter and the secondary receiver, which is assumed to be a zero-mean Gaussian random process with power spectral density $S_{h}(e^{jw})$. It is further assumed that $x_{k}$ is independent of $h_{k}$ and $n_{k}$. While both the secondary transmitter and the secondary receiver know the channel statistics, neither has prior knowledge of instantaneous realizations of the fading coefficients. In (\ref{input-out1}), $s_{p,k}$ represents the sum of the active primary users' faded signals arriving at the secondary receiver. We assume that the bandwidth available in the system is $B$, and the channel is sensed every first $N$ seconds of the block which is $T$ seconds. Note that the discrete-time model is obtained by sampling the received signal every $T_{s}=\frac{1}{B}$ seconds.

If the transmission strategies of the primary users are not known, energy-based detection methods are well-suited for the detection of the activities of primary users. Following the same approach as in \cite{globecom_akin}, we consider a hypothesis testing problem between the noise $n_{k}$ and the signal $s_{p,k}$ in noise. Noting
that there are $NB$ complex symbols in a duration of $N$ seconds, this can mathematically be expressed as follows:
\begin{align}\label{hypothesis}
&\mathcal{H}_{0}\quad : \quad y_{k}=n_{k}, \quad k=1,\dots,NB\\ \nonumber
&\mathcal{H}_{1}\quad : \quad y_{k}=s_{p,k}+n_{k}, \quad k=1,\dots,NB.
\end{align}
Considering the above detection problem, the optimal Neyman-Pearson detector is given by \cite{Poor-book} as
\begin{equation}\label{Neyman-Pearson}
Y=\frac{1}{NB}\sum_{k=0}^{NB-1}|y_{k}|^{2}\gtrless^{\mathcal{H}_{1}}_{\mathcal{H}_{0}}\lambda
\end{equation}
where $\lambda$ is the detection threshold. We assume that $s_{p,k}$ has a circularly symmetric complex Gaussian distribution with zero-mean and variance $\sigma_{s_{p}}^{2}$. Assuming further that $\{s_{p,k}\}$ are i.i.d., we can immediately conclude that the test statistic $Y$ is chi-square distributed with $2NB$ degrees of freedom. In this case, the probabilities of false alarm and detection can be established as follows:
\begin{align}\label{false alarm}
&P_{f}=Pr(Y>\lambda|\mathcal{H}_{0})=1-P\left(\frac{NB\lambda}{\sigma_{n}^{2}},NB\right)\\
&P_{d}=Pr(Y>\lambda|\mathcal{H}_{1})=1-P\left(\frac{NB\lambda}{\sigma_{n}^{2}+\sigma_{s_{p}}^{2}},NB\right) \label{eq:probdetect}
\end{align}
where $P(x,a)$ denotes the regularized lower gamma function and is defined as $P(x,a) = \frac{\gamma(x,a)}{\Gamma(a)}$ where $\gamma(x,a)$ is the lower incomplete gamma function and $\Gamma(a)$ is the Gamma function.

In the above hypothesis testing problem, another approach is to consider $Y$ as Gaussian distributed, which is accurate if $NB$ is large \cite{Hoang}. In this case, the detection and false alarm probabilities can be expressed in terms of Gaussian $Q$-functions. We would like to note the rest of the analysis in the paper does not depend on the specific expressions of the false alarm and detection probabilities. However, numerical results are obtained using (\ref{false alarm}) and (\ref{eq:probdetect}).

\vspace{-0.2cm}
\section{Pilot Symbol Assisted Modulation and Channel Estimation at the Receiver}\label{PSAM}
We consider pilot-assisted transmission where periodically inserted pilot symbols, known by both the sender and the receiver, are used to estimate the fading coefficients of the channel. We assume a simple scenario where a pilot symbol is transmitted every $T$ seconds after channel sensing, and $(T-N)B-1$ data symbols following the pilot symbol are transmitted between the secondary transmitter and the secondary receiver. We consider the following average power constraints:
\begin{equation}\label{power constraint 1}
\frac{1}{T}\sum_{k=lTB+NB}^{(l+1)TB-1}|x_{1,k}|^{2}\leq \overline{P}_{1} \quad l=\dots,-2,-1,0,1,2,\dots
\end{equation}
when the channel is detected as busy. On the other hand, when the channel is detected as idle the power constraint becomes
\begin{equation}\label{power constraint 1}
\frac{1}{T}\sum_{k=lTB+NB}^{(l+1)TB-1}|x_{2,k}|^{2}\leq \overline{P}_{2} \quad l=\dots,-2,-1,0,1,2,\dots
\end{equation}
Therefore, the total number of symbols transmitted in one block is $(T-N)B$, and the total energy is limited by $T\overline{P}_{1}$ and $T\overline{P}_{2}$ for the cases when the channel is busy and idle, respectively.

Communication takes place in two phases. In the training phase, the transmitter sends pilot symbols and the receiver estimates the channel coefficients. In this phase, the channel output is given by
\begin{equation}\label{training1}
y_{lTB+NB}=h_{lTB+NB}\sqrt{P_{t}}+n_{lTB+NB}+s_{p,lTB+NB}
\end{equation}
when the channel is busy, and
\begin{equation}\label{training2}
y_{lTB+NB}=h_{lTB+NB}\sqrt{P_{t}}+n_{lTB+NB}
\end{equation}
when the channel is idle. $P_{t}$ is the power allocated to the pilot symbol. In the data transmission phase, data symbols are transmitted. In this phase, input-output relationship can be written as
\begin{equation}\label{datatransmission1}
y_{k}=\widehat{h}_{k}x_{1,k}+\widetilde{h}_{k}x_{1,k}+n_{k}+s_{p,k}\quad lTB+NB<k\leq (l+1)TB-1
\end{equation}
when the channel is busy, and
\begin{equation}\label{datatransmission11}
y_{k}=\widehat{h}_{k}x_{2,k}+\widetilde{h}_{k}x_{2,k}+n_{k}\quad lTB+NB<k\leq (l+1)TB-1
\end{equation}
when the channel is idle where $\widehat{h}_{k}$ and $\widetilde{h}_{k}$ are the estimated channel coefficient and the error in the estimate at sample time $k$, respectively. Note that $\widehat{h}_{k}$ and $\widetilde{h}_{k}$ for $lTB+NB<k\leq (l+1)TB-1$ are uncorrelated zero-mean circularly symmetric complex Gaussian random variables with variances $\sigma_{\widehat{h}_{k}}^{2}$ and $\sigma_{\widetilde{h}_{k}}^{2}$, respectively.

Let us assume that $\mathcal{C}$ is the set of integers, $l$, such that outputs at time samples $lTB+NB$ are used to estimate the channel fading coefficient at time $k$. Due to periodicity, we will consider only the cases where $k=NB,NB+1,\dots,TB-1$. We consider a causal and a non-causal estimation procedure.

\section{Achievable Rates}
Considering the decision of channel sensing and its correctness, we have four possible scenarios:
\begin{enumerate}
  \item Channel is busy, detected as busy,
  \item Channel is busy, detected as idle,
  \item Channel is idle, detected as busy,
  \item Channel is idle, detected as idle.
\end{enumerate}

We assume that a Wiener filter, known as the optimum linear estimator in the mean-square sense, is used at the secondary receiver to estimate the fading coefficients of the channel between the secondary transmitter and the secondary receiver. Since the pilot symbol is sent every $T$ seconds, the channel is sampled every $T$ seconds. Therefore, we have to consider the under-sampled version of the channel's Doppler spectrum given by
\begin{equation}\label{under-sampled}
S_{h,m}(e^{jw})=\frac{1}{TB}\sum_{i=0}^{TB-1}e^{\frac{jm(w-2i\pi}{TB})}S_{h}(e^{\frac{j(w-2i\pi)}{TB}}).
\end{equation}
Also as shown in \cite{spawc_akin}, \cite{Hassibi} and \cite{giannakis}, it can be easily obtained that the channel MMSE for the noncausal Wiener filter at time $lTB+NB+m$ is given by
\begin{equation}\label{noncausal min mean square error}
\sigma_{\widetilde{h}_{TBl+NB+m}}^{2}=\sigma_{h}^{2}-\frac{P_{t}}{2\pi}\int_{-\pi}^{\pi}\frac{|S_{h,m}(e^{jw})|^{2}}{P_{t}S_{h,0}(e^{jw})+\sigma_{n}^{2}+\rho\sigma_{s_{p}}^{2}}\mathit{dw}.
\end{equation}
where $\rho$ is the activity probability of primary users in the channel. As for the causal Wiener filter, using a similar approach as in in \cite{spawc_akin} and \cite{Hassibi}, MMSE at time $lTB+NB+m$ is given by
\begin{align}\label{causal min mean square error}
\sigma_{\widetilde{h}_{lTB+NB+m}}^{2}=\sigma_{h}^{2}-&\frac{P_{t}}{2\pi}\int_{-\pi}^{\pi}\frac{|S_{h,m}(e^{jw})|^{2}}{P_{t}S_{h,0}(e^{jw})+\sigma_{n}^{2}+\rho\sigma_{s_{p}}^{2}}\mathit{dw}\nonumber\\+&\frac{1}{2\pi}\int_{-\pi}^{\pi}\frac{P_{t}}{r_{e}}\left|\left\{\frac{S_{h,m}(e^{jw})}{L^{*}(e^{jw})}\right\}_{-}\right|^{2}\mathit{dw}.
\end{align}
where $L^{*}(e^{jw})$ is obtained from the canonical factorization of the channel output's sampled power spectral density at $m=0$, given by
\begin{equation}\label{canon}
P_{t}S_{h,0}(e^{jw})+\sigma_{n}^{2}+\rho\sigma_{s_{p}}^{2}=r_{e}L(e^{jw})L^{*}(e^{jw}).
\end{equation}
The operators $\{\}_{+}$ and $\{\}_{-}$ give the causal and anti-causal parts of the function to which they are applied, respectively. Using the orthogonality principle we have
\begin{equation}\label{orhogonal}
\sigma_{\widehat{h}_{lTB+NB+m}}^{2}=\sigma_{h}^{2}-\sigma_{\widetilde{h}_{lTB+TB+m}}^{2}
\end{equation}
where $\sigma_{\widehat{h}_{lTB+NB+m}}^{2}$ is the variance of the channel estimate at time $lTB+NB+m$. Similarly as in \cite{spawc_akin}, we regard the errors in (\ref{datatransmission1}) and (\ref{datatransmission11}) as another source of additive noise and assume that
\begin{equation}\label{new noise 1}
w_{1,k}=\widetilde{h}_{k}x_{1,k}+n_{k}+s_{p,k}
\end{equation}
and
\begin{equation}\label{new noise 2}
w_{2,k}=\widetilde{h}_{k}x_{2,k}+n_{k}
\end{equation}
are zero-mean Gaussian noises with variances
\begin{equation}\label{new variance 1}
\sigma_{w_{1,k}}^{2}=\sigma_{\widetilde{h}_{k}}^{2}P_{1,m}+\sigma_{n}^{2}+\sigma_{s_{p,k}}^{2}
\end{equation}
when the channel is busy, and
\begin{equation}\label{new variance 2}
\sigma_{w_{2,k}}^{2}=\sigma_{\widetilde{h}_{k}}^{2}P_{2,m}+\sigma_{n}^{2}
\end{equation}
when the channel is idle, respectively. Considering the errors in (\ref{datatransmission1}) and (\ref{datatransmission11}) and the scenarios given above, we obtain the following lower bound on the channel capacity:

\begin{align}\label{capacity_gh}
&C\geq \frac{1}{T}\sum_{m=1}^{(T-N)B-1}E\bigg\{\rho P_{d}\log\left(1+\frac{P_{1,m}\sigma_{\widehat{h}_{m}}^{2}}{P_{1,m}\sigma_{\widetilde{h}_{m}}^{2}+\sigma_{n}^{2}+\sigma_{s_{p}}^{2}}|\xi|^{2}\right)\nonumber\\
&+\rho(1-P_{d})\log\left(1+\frac{P_{2,m}\sigma_{\widehat{h}_{m}}^{2}}{P_{2,m}\sigma_{\widetilde{h}_{m}}^{2}+\sigma_{n}^{2}+\sigma_{s_{p}}^{2}}|\xi|^{2}\right)\nonumber\\ &+(1-\rho)P_{f}\log\left(1+\frac{P_{1,m}\sigma_{\widehat{h}_{m}}^{2}}{P_{1,m}\sigma_{\widetilde{h}_{m}}^{2}+\sigma_{n}^{2}}|\xi|^{2}\right)\nonumber\\
&+(1-\rho)(1-P_{f})\log\left(1+\frac{P_{2,m}\sigma_{\widehat{h}_{m}}^{2}}{P_{2,m}\sigma_{\widetilde{h}_{m}}^{2}+\sigma_{n}^{2}}|\xi|^{2}\right)
\bigg\}
\end{align}
where $\xi$ is a zero-mean, unit variance, circularly symmetric complex Gaussian random variable, and $P_{1,m}=E\left[|x_{1,lTB+NB+m}|^{2}\right]$ and $P_{2,m}=E\left[|x_{2,lTB+NB+m}|^{2}\right]$ denote the power of $m^{th}$ data symbol after the pilot symbol when the channel is busy and idle, respectively. Note that the error variance $\sigma_{\widetilde{h}_{lTB+NB+m}}^{2}$ depends on $m$ or the location of data symbol with respect to the pilot symbol. However, if the fading is slowly varying and the channel sampled fast enough, we can satisfy $f_{D}\leq\frac{1}{2T}$ or $w_{D}\leq\frac{\pi}{BT}$ where $f_{D}$ is the maximum Doppler frequency and $Bw_{D}=2\pi f_{D}$. In this case, we can see from Nyquist's Theorem that there is no aliasing in the under-sampled version of the channel's Doppler spectrum, and hence $|S_{h,m}(e^{jw})|=|S_{h,0}(e^{jw})|=S_{h}(e^{jw/T/B})/T/B$, for $m\in[1,TB]$ and $-\pi\leq w\leq\pi$. Henceforth, (\ref{noncausal min mean square error}) becomes
\begin{align}\label{noncausal_min mean square error new}
&\sigma_{\widetilde{h}_{lTB+NB+m}}^{2}=\sigma_{h}^{2}-\frac{P_{t}}{2\pi}\int_{-\pi}^{\pi}\frac{|S_{h,m}(e^{jw})|^{2}}{P_{t}S_{h,0}(e^{jw})+\sigma_{n}^{2}+\rho\sigma_{s_{p}}^{2}}\mathit{dw}\nonumber\\ &=\sigma_{h}^{2}-\frac{P_{t}}{2\pi}\int_{-\frac{\pi}{TB}}^{\frac{\pi}{TB}}\frac{|S_{h}(e^{jw})|^{2}}{P_{t}S_{h}(e^{jw})+TB(\sigma_{n}^{2}+\rho\sigma_{s_{p}}^{2})}\mathit{dw}=\sigma_{\widetilde{h}}^{2}
\end{align}
and (\ref{causal min mean square error}) becomes
\begin{align}\label{causal min mean square error new}
&\sigma_{\widetilde{h}_{lTB+NB+m}}^{2}=\sigma_{h}^{2}-\frac{P_{t}}{2\pi}\int_{-\pi}^{\pi}\bigg[\frac{|S_{h,m}(e^{jw})|^{2}}{P_{t}S_{h,0}(e^{jw})+\sigma_{n}^{2}+\rho\sigma_{s_{p}}^{2}}\nonumber\\ &-\frac{1}{r_{e}}\left|\left\{\frac{S_{h,m}(e^{jw})}{L^{*}(e^{jw})}\right\}_{-}\right|^{2}\bigg]\mathit{dw}\nonumber\\
&=\sigma_{h}^{2}-\frac{P_{t}}{2\pi}\int_{-\frac{\pi}{TB}}^{\frac{\pi}{TB}}\frac{|S_{h}(e^{jw})|^{2}}{P_{t}S_{h}(e^{jw})+TB(\sigma_{n}^{2}+\rho\sigma_{s_{p}}^{2})}\mathit{dw}\nonumber\\ &+\frac{P_{t}}{2\pi r_{e}(TB)^{2}}\int_{-\pi}^{\pi}\left|\left\{\frac{S_{h}(e^{\frac{jw}{TB}})}{L^{*}(e^{jw})}\right\}_{-}\right|^{2}\mathit{dw}=\sigma_{\widetilde{h}}^{2},
\end{align}
where
\begin{equation}\label{new F}
\frac{P_{t}S_{h}(e^{\frac{jw}{TB}})}{TB}+\sigma_{n}^{2}+\rho\sigma_{s_{p}}^{2}=r_{e}L(e^{jw})L^{*}(e^{jw}).
\end{equation}
Therefore, under this assumption, the error variances become independent of $m$. Since the estimate quality is the same for each data symbol regardless of its position with respect to the pilot symbol, uniform power allocation among the data symbols is optimal, and we have
\begin{equation}\nonumber
P_{1,m}=\frac{T\overline{P}_{1}-P_{t}}{(T-N)B-1}=P_{1}
\end{equation}
and
\begin{equation}\nonumber
P_{2,m}=\frac{T\overline{P}_{2}-P_{t}}{(T-N)B-1}=P_{2}.
\end{equation}
Then, (\ref{capacity_gh}) becomes
\begin{align}\label{capacity_ghh}
&C\geq \frac{(T-N)B-1}{T}E\bigg\{\rho P_{d}\log\left(1+\frac{P_{1}\sigma_{\widehat{h}}^{2}}{P_{1}\sigma_{\widetilde{h}}^{2}+\sigma_{n}^{2}+\sigma_{s_{p}}^{2}}|\xi|^{2}\right)\nonumber\\
&+\rho(1-P_{d})\log\left(1+\frac{P_{2}\sigma_{\widehat{h}}^{2}}{P_{2}\sigma_{\widetilde{h}}^{2}+\sigma_{n}^{2}+\sigma_{s_{p}}^{2}}|\xi|^{2}\right)\nonumber\\ &+(1-\rho)P_{f}\log\left(1+\frac{P_{1}\sigma_{\widehat{h}}^{2}}{P_{1}\sigma_{\widetilde{h}}^{2}+\sigma_{n}^{2}}|\xi|^{2}\right)\nonumber\\
&+(1-\rho)(1-P_{f})\log\left(1+\frac{P_{2}\sigma_{\widehat{h}}^{2}}{P_{2}\sigma_{\widetilde{h}}^{2}+\sigma_{n}^{2}}|\xi|^{2}\right)
\bigg\}
\end{align}

\section{Numerical Results in Gauss-Markov Channels}
In this section, we consider Gauss-Markov fading process whose dynamics is given as
\begin{equation}\label{autocorrelation}
h_{k}=\alpha h_{k-1}+z_{k}\quad 0\leq \alpha\leq1\quad k=1,2,3,...
\end{equation}
where $\alpha$ is a parameter that controls the rate of channel variation, and $\{z_{k}\}$ are i.i.d. circular complex Gaussian variables with zero mean and variance $(1-\alpha)\sigma_{h}^{2}$. The power spectral density of this channel is given as
\begin{equation}\label{psd}
S_{h}(e^{jw})=\frac{(1-\alpha^{2})\sigma_{h}^{2}}{1+\alpha^{2}-2\alpha\cos(w)}\quad -\pi\leq w \leq\pi.
\end{equation}
Note that $S_{h}(e^{jw})$ in (\ref{psd}) is not band-limited. Therefore, the condition $w_{D}\leq\frac{\pi}{BT}$ can be satisfied only when $TB=1$ which is not a reliable strategy. On the other hand, if the fading is varying slowly and the value of $\alpha$ is close to 1, the Doppler spectrum $S_{h}(e^{jw})$ decreases sharply for large frequencies and most of the energy is kept around the lower frequencies. We can easily find that the frequency ranges $[-\pi/49,\pi/49]$, $[-\pi/9,\pi/9]$ and $[-\pi/4,\pi/4]$ contain more than $90\%$ of the power when $\alpha=0.99,0.95$ and $0.90$, respectively. Therefore, in these cases the effect of aliasing is negligible.

We can easily obtain the error variance for the noncausal Wiener filter from (\ref{noncausal_min mean square error new}). On the other hand, we have to perform canonical factorization in order to obtain the error variance for the causal Wiener filter in the absence of aliasing. Due to limited space in the paper, we will skip some of the calculation steps. We can express the anti-causal part of (\ref{causal min mean square error new}) as
\begin{equation}\label{anti-causal part}
\left\{\frac{S_{h}\left(e^{\frac{jw}{TB}}\right)}{L^{*}\left(e^{jw}\right)}\right\}_{-}=\frac{\left(1-\alpha^{2}\right)\sigma_{h}^{2}v}{1-\alpha v}\frac{e^{\frac{jw}{TB}}}{\left(1-ve^{\frac{jw}{TB}}\right)}.
\end{equation}
where
\begin{equation}\label{v}
v=\frac{\alpha\left(\sigma_{n}^{2}+\rho\sigma_{s_{p}}^{2}\right)}{r_{e}}
\end{equation}
\begin{equation}\label{re}
r_{e}=\frac{c+\sqrt{c^{2}-4\alpha^{2}\left(\sigma_{n}^{2}+\rho\sigma_{s_{p}}^{2}\right)^{2}}}{2}
\end{equation}
and
\begin{equation}\label{c}
c=\frac{P_{t}}{TB}\left(1-\alpha^{2}\right)\sigma_{h}^{2}+\left(1+\alpha^{2}\right)\left(\sigma_{n}^{2}+\sigma_{s_{p}}^{2}\right)
\end{equation}
Note that
\begin{equation}\label{equation}
L\left(e^{jw}\right)=\frac{1-ve^{-\frac{jw}{TB}}}{1-\alpha e^{-\frac{jw}{TB}}}.
\end{equation}
Using (\ref{anti-causal part}) and (\ref{re}), we can obtain the error variance for the causal Wiener filter as
\begin{align}\label{causal min mean square error en son}
\sigma_{\widetilde{h}}^{2}&=\sigma_{h}^{2}-\frac{P_{t}}{2\pi}\int_{-\frac{\pi}{TB}}^{\frac{\pi}{TB}}\bigg[\frac{|S_{h}(e^{jw})|^{2}}{P_{t}S_{h}(e^{jw})+TB(\sigma_{n}^{2}+\rho\sigma_{s_{p}}^{2})}\nonumber\\ &-\frac{(1-\alpha^{2})^{2}\sigma_{h}^{4}v^{2}}{TBr_{e}(1-\alpha v)^{2}(1+v^{2}-2v\cos(w))}\bigg]\mathit{dw}
\end{align}

Since the primary users are not to be disturbed by the secondary users, when the primary users are active there is an average transmission power threshold on the secondary users which is denoted as $I_{avg}$. Therefore, the power constraint can be expressed as follows
\begin{equation}\label{interference}
P_{t}+\left[(T-N)B-1\right]\left[P_{d}P_{1}+(1-P_{d})P_{2}\right]\leq TI_{avg}
\end{equation}
and the average signal-to-noise (SNR) is
\begin{equation}\label{snr}
SNR = \frac{I_{avg}}{B\sigma_{n}^{2}}.
\end{equation}

\begin{figure}
\begin{center}
\includegraphics[width = \figsize\textwidth]{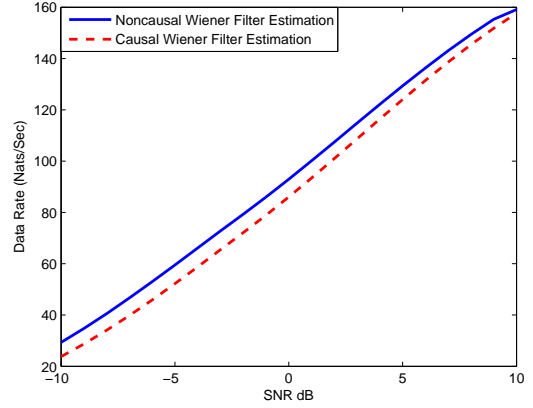}
\caption{SNR dB v.s. Data Rate (Nats/Sec) when $B=100$Hz, $T=0.5$sec, $N=0.1$sec, $\alpha=0.99$, $P_{d}=0.91$, and $P_{f}=0.23$.} \label{fig:fig1}
\end{center}
\end{figure}

\begin{figure}
\begin{center}
\includegraphics[width = \figsize\textwidth]{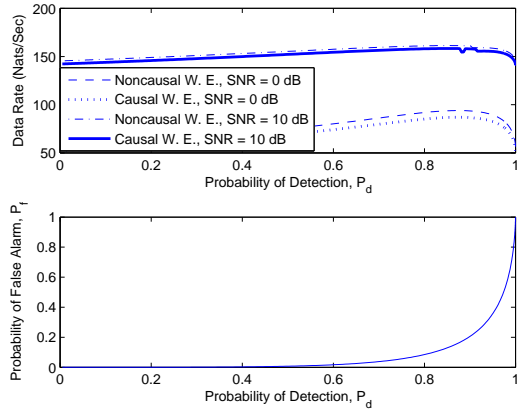}
\caption{Data Rate and Probability of False Alarm, $P_{f}$ v.s. Probability of Detection, $P_{d}$ when $B=100$Hz, $T=0.5$sec, $N=0.1$sec, and $\alpha=0.99$.} \label{fig:fig2}
\end{center}
\end{figure}

\begin{figure}
\begin{center}
\includegraphics[width =\figsize\textwidth]{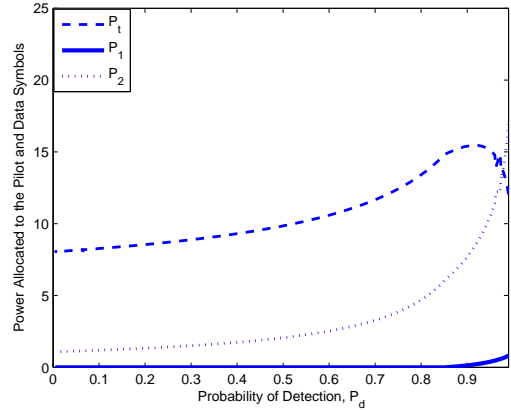}
\caption{The optimal power distribution among the pilot and data symbols when $B=100$Hz, $T=0.5$sec, $N=0.1$sec, and $\alpha=0.99$.} \label{fig:fig3}
\end{center}
\end{figure}

In Figure \ref{fig:fig1}, employing a causal Wiener filter and a noncausal Wiener filter at the secondary receiver, we show the data rate as a function of SNR (dB) when $B=100$Hz, and $T=0.5$ seconds and $N=0.1$ seconds. We choose $P_{d}=0.91$ and $P_{f}=0.23$ since in reality these are plausible and obtainable values considering the tradeoff between probability of detection and false alarm. Note that when $P_{d}=1$, there will be no constraint on $P_{2}$ and the secondary user can use power as much as its peak value allows. As expected, we obtain higher data rates when we use a noncausal Wiener filter. Note that the difference between the data rates obtained by using a causal and a noncausal Wiener filter is decreasing with the increasing SNR. In Fig. \ref{fig:fig2}, we compare the data rate values for different SNR values, 0 dB and 10 dB, as a function of probability of detection, $P_{d}$, again employing a causal and a noncausal Wiener filter at the secondary receiver. It is clearly seen that the highest data rates are obtained when $P_{d}=0.91$ and $P_{f}=0.23$ for both Wiener filters. Again, as expected the data rates obtained using noncausal Wiener filters are higher than the ones obtained by causal Wiener filters. In Fig. \ref{fig:fig3}, we display the optimal power distribution as a function of probability of detection, $P_{d}$. Recall that $P_{t}$ denotes the power allocated to the pilot symbol, and $P_{1}$ and $P_{2}$ represent the power allocated to the data symbols when the channel is busy and idle, respectively. In Fig. \ref{fig:fig3}, it is easily observed that the power allocated to the data symbols when the channel is busy or idle is decreasing with decreasing $P_{d}$, since the interference power constraint is mainly over $P_{2}$ and there is less transmission with $P_{1}$. Furthermore, the power allocated to the pilot symbol is the highest when $P_{d}=0.91$, where at the same time $P_{f}=0.23$.

\section{Conclusion}
We have studied channel sensing, pilot-symbol assisted channel estimation, and achievable rates when causal and noncausal Wiener filters are employed at the secondary receiver. We have obtained achievable rates by finding error variances in both cases, and shown the effects of channel sensing over the achievable rates. Subsequently, we have optimized power allocated to the pilot and data symbols. Finally, we have obtained numerical results in Gauss-Markov channels showing the optimal parameter values. We have compared the performances of causal and noncausal Wiener filters at different SNR, $P_{d}$ and $P_{f}$ values.


\begin{thebibliography}{99}

\bibitem{cognitive} ``Cognitive Radio Technology,"
\emph{IEEE Signal Process. Mag., Nov. 2008.}

\bibitem{qzhao} Q. Zhao, S. Geirhofer, L. Tong, and B. M. Sadler, ``Opportunistic Spectrum Access via Periodic Channel Sensing,"
\emph{IEEE Trans. on Signal Processing, vol. 56, no. 2, pp. 785-796, Feb. 2008.}

\bibitem{yunxia} Y. Chen, Q. Zhao, and A. Swami, ``Joint Design and Seperation Principle for Opportunistic Spectrum Access,"
\emph{in Proc. 40th IEEE Asilomar Conf. Signals, Systems, Comput., Oct. 2006, pp. 696 - 700.}

\bibitem{Hoang} Y.-C. Liang, Y. Zheng, E. C. Y. Peh, and A. T. Hoang, ``Sensing-throughput tradeoff for cognitive radio networks,"
\emph{IEEE Trans. Wireless Commun., vol. 7, no. 4, pp. 1326-1337, Apr. 2008.}

\bibitem{poor} Z. Quan, S. Cui, A. H. Sayed, and H. V. Poor,``Wideband Spectrum Sensing in Cognitive Radio Networks,"
\emph{Proc. of IEEE International Conference on Communications, Beijing, China, May 19-23, 2008.}

\bibitem{cavers1} J.K. Cavers, ``An analysis of pilot symbol assisted modulation for Rayleigh fading channels,"
\emph{IEEE Trans. Vehicular Tech., vol. 40, pp. 686-693, November 1991.}

\bibitem{cavers2} 1.K. Cavers, ``Pilot assisted symbol modulation and differential detection in fading and delay spread,"
\emph{IEEE Trans. Inform. Theory, vol. 43, no. 7, pp. 2206-2212, 1995.}

\bibitem{hassibi} B. Hassibi, and B. M. Hochwald, ``How much training is needed in multiple-antenna wireless links?,"
\emph{IEEE Trans. Inform. Theory, Vol. 49, pp. 951-963, April 2003.}

\bibitem{giannakis} S. Ohno and G.B. Giannakis, ``Average-Rate Optimal PSAM Transmissions Over Time-Selective Fading Channels,"
\emph{IEEE Trans. on Wireless Comm., Vol. 1, No.4, October 2002.}

\bibitem{faycal} I. Abou-Faycal, M. M$\acute{e}$dard, and U. Madhow, ``Binary adaptive coded pilot symbol assisted modulation over Rayleigh fading channels without feedback,"
\emph{IEEE Trans. Commun., vol. 53, pp. 1036-1046, June 2005.}

\bibitem{icc_akin} S. Akin and M. C. Gursoy, ``Training Optimization for Gauss-Markov Rayleigh Fading Channels,"
\emph{Proceedings ofthe 2007 IEEE International Conference on Communications, Glasgow, Scottland, UK, 2428 June 2007.}

\bibitem{spawc_akin} S. Akin and M. C. Gursoy, ``Pilot-Symbol-Assisted Communications with Noncausal and Causal Wiener Filters,"
\emph{Proc. of the 9th IEEE International Workshop on Signal Processing Advances in Wireless Communications (SPAWC), Recife, Brazil, July 2008.}

\bibitem{globecom_akin} S. Akin and M. C. Gursoy, ``Effective Capacity Analysis of Cognitive Radio Channels for Quality of Service Provisioning,"
\emph{IEEE Global Communication Conference, Honolulu, Hawaii, Nov. 30 - Dec. 4, 2009.}

\bibitem{Poor-book} H. V. Poor, ``An Introduction to Signal Detection and Estimation, 2nd ed.,"
\emph{Springer-Verlag, 1994.}

\bibitem{Hassibi} T. Kailath, A.H. Sayed, and B. Hassibi, ``Linear Estimation,"
\emph{Upper Saddle River, New Jersey: Prentice Hall, 2000.}

\end{thebibliography}
\end{document}